\documentclass[%
twocolumn,
longbibliography,
superscriptaddress,
amsmath,amssymb,
aps,
prb,
]{revtex4-1}

\usepackage{graphicx}
\usepackage{dcolumn}
\usepackage{bm}
\usepackage{mciteplus}
\usepackage[version=3]{mhchem}
\usepackage{color}

\graphicspath{{figs/}}

\begin{document}
	\title{Optical harmonic generation in monolayer group-VI  transition metal dichalcogenides}

\author{Anton Autere}
\affiliation
{Department of Electronics and Nanoengineering, Aalto University, P.O.Box 13500, FI-00076 Aalto, Finland}

\author{Henri Jussila}
\affiliation
{Department of Electronics and Nanoengineering, Aalto University, P.O.Box 13500, FI-00076 Aalto, Finland}

\author{Andrea Marini}
\affiliation
{ICFO-Institut de Ci{\`e}ncies Fot{\`o}niques, The Barcelona Institute of Science and Technology, 08860 Castelldefels, Barcelona, Spain}

\author{J. R. M. Saavedra}
\affiliation
{ICFO-Institut de Ci{\`e}ncies Fot{\`o}niques, The Barcelona Institute of Science and Technology, 08860 Castelldefels, Barcelona, Spain}

\author{Yunyun Dai}
\affiliation
{Department of Electronics and Nanoengineering, Aalto University, P.O.Box 13500, FI-00076 Aalto, Finland}

\author{Antti S{\"a}yn{\"a}tjoki}
\affiliation
{Department of Electronics and Nanoengineering, Aalto University, P.O.Box 13500, FI-00076 Aalto, Finland}
\affiliation
{Institute of Photonics, University of Eastern Finland, Yliopistokatu 7, FI-80101 Joensuu, Finland}

\author{Lasse Karvonen}
\affiliation
{Department of Electronics and Nanoengineering, Aalto University, P.O.Box 13500, FI-00076 Aalto, Finland}

\author{He Yang}
\affiliation
{Department of Electronics and Nanoengineering, Aalto University, P.O.Box 13500, FI-00076 Aalto, Finland}

\author{Babak Amirsolaimani}
\affiliation
{College of Optical Sciences, University of Arizona, 1630 E. University Boulevard, Tucson, Arizona 85721, USA}

\author{Robert A. Norwood}
\affiliation{College of Optical Sciences, University of Arizona, 1630 E. University Boulevard, Tucson, Arizona 85721, USA}

\author{Nasser Peyghambarian}
\affiliation{College of Optical Sciences, University of Arizona, 1630 E. University Boulevard, Tucson, Arizona 85721, USA}
\affiliation
{Institute of Photonics, University of Easter\textbf{}n Finland, Yliopistokatu 7, FI-80101 Joensuu,
	Finland}
\affiliation
{Department of Electronics and Nanoengineering, Aalto University, P.O.Box 13500, FI-00076 Aalto, Finland}

\author{Harri Lipsanen}
\affiliation
{Department of Electronics and Nanoengineering, Aalto University, P.O.Box 13500, FI-00076 Aalto, Finland}

\author{Khanh Kieu}
\affiliation
{College of Optical Sciences, University of Arizona, 1630 E. University Boulevard, Tucson, Arizona 85721, USA}

\author{F.~Javier~Garc\'{\i}a~de~Abajo}
\affiliation
{ICFO-Institut de Ci{\`e}ncies Fot{\`o}niques, The Barcelona Institute of Science and Technology, 08860 Castelldefels, Barcelona, Spain}
\affiliation
{ICREA-Instituci\'o Catalana de Recerca i Estudis Avan\c{c}ats, Passeig Llu\'{\i}s Companys 23, 08010 Barcelona, Spain}

\author{Zhipei Sun}
\affiliation
{Department of Electronics and Nanoengineering, Aalto University, P.O.Box 13500, FI-00076 Aalto, Finland}
\affiliation
{QTF Centre of Excellence, Department of Applied Physics, Aalto University, FI-00076 Aalto, Finland}
\email{zhipei.sun@aalto.fi}

\newcommand{\smos}{5.4}
\newcommand{\smose}{37.0}
\newcommand{\sws}{16.2}
\newcommand{\swse}{16.5}

\newcommand{\tmos}{3.6}
\newcommand{\tmose}{2.2}
\newcommand{\tws}{2.4}
\newcommand{\twse}{1.0}
\begin{abstract}
Monolayer transition metal dichalcogenides (TMDs) exhibit high nonlinear optical (NLO) susceptibilities. Experiments on MoS$_2$ have indeed revealed very large second-order ($\chi^{(2)}$) and third-order ($\chi^{(3)}$) optical susceptibilities. However, third harmonic generation results of other layered TMDs has not been reported. Furthermore, the reported $\chi^{(2)}$ and $\chi^{(3)}$ of MoS$_2$ vary by several orders of magnitude, and a reliable quantitative comparison of optical nonlinearities across different TMDs has remained elusive. Here, we investigate second- and third-harmonic generation, and three-photon photoluminescence in TMDs. Specifically, we present an experimental study of $\chi^{(2)}$, and $\chi^{(3)}$ of four common TMD materials (\ce{MoS2}, \ce{MoSe2}, \ce{WS2} and \ce{WSe2}) by placing different TMD flakes in close proximity to each other on a common substrate, allowing their NLO properties to be accurately obtained from a single measurement. $\chi^{(2)}$ and $\chi^{(3)}$ of the four monolayer TMDs have been compared, indicating that they exhibit distinct NLO responses. We further present theoretical simulations of these susceptibilities in qualitative agreement with the measurements. Our comparative studies of the NLO responses of different two-dimensional layered materials allow us to select the best candidates for atomic-scale nonlinear photonic applications, such as frequency conversion and all-optical signal processing.
	
\end{abstract}
\maketitle
\section{Introduction}
Recent years have witnessed a growing interest in two-dimensional (2D) layered materials for various electronic and photonic applications\cite{Ferrari_N_2015}. This includes graphene\cite{Bonaccorso_NP_2010} and transition metal dichalcogenides (TMDs), especially molybdenum disulfide (MoS$_2$)\cite{mak2010,splendiani2010,wang2012}. TMDs possess fascinating layer-dependent optical and electrical properties, such as their layer-dependent band gap. For example, bulk (group-VI) TMDs are typically indirect band-gap semiconductors, while in single atomic layers their band gap becomes direct in the $\sim$1.55 eV-1.9 eV range \cite{jin2013direct,mak2010,splendiani2010}. This renders monolayer TMDs (ML-TMDs) as attractive materials for various optoelectronic applications, such as light-emitting devices, detectors and modulators \cite{sun2016,wang2012}. ML-TMDs consist of two hexagonal lattices of chalcogen atoms separated by a plane of metal atoms occupying trigonal prismatic sites between the chalcogens \cite{wang2012}. Owing to their crystal structure, TMDs with an odd number of layers are noncentrosymmetric, while TMDs in bulk or with any even number of layers are centrosymmetric\cite{li2013}. The lack of inversion symmetry in ML-TMDs leads to substantial second-order nonlinear optical susceptibility $\chi^{(2)}$.

Nonlinear optical (NLO) processes in 2D materials are of great interest for various technological applications\cite{Autere_AM_18,gu2012regenerative,hendry2010coherent,Li2018,Yang2018,Sun2018}, such as frequency conversion, all-optical signal processing, ultrafast pulse generation, and parametric sources of quantum photon states. Furthermore, integration of 2D materials with photonic integrated circuits offers exciting prospects for new applications. In particular, it has already been shown that the NLO responses of 2D materials can be enhanced with waveguides\cite{Chen_LSA_2017,Autere_AM_18} and photonic crystal structures\cite{Autere_AM_18,Taylor_2M_2017,gu2012regenerative}. The NLO properties of 2D materials are promising for the development of on-chip devices, such as nonlinear light sources for quantum photonics and metrology or nonlinear phase modulation devices\cite{Autere_AM_18,wang2015tunable}. In addition, the fundamental properties (e.g. crystal orientation) of different 2D materials can be probed via nonlinear optical processes such as second-harmonic generation (SHG), which is important for nanomaterial characterization \cite{Autere_AM_18,autere2017rapid,liang2017monitoring,zhou2015strong}. Thus far, research on TMDs has been focused on their electronic and linear-optical properties, with only few studies reporting on NLO properties.  Several groups have already reported the observation of SHG in mono- and trilayer MoS$_2$ \cite{malard2013,li2013,kumar2013second,Lin_NaLe_18}, as well as in \ce{MoSe2}\cite{le2016nonlinear}, WS$_2$\cite{janisch2014}, and WSe$_2$ \cite{zeng2013optical}. Additionally, third order optical nonlinearity, quantified through the third-order susceptibility $\chi^{(3)}$, has been recently observed in few-layer\cite{wang2013third} and monolayer\cite{antti2016,woodward2016characterization,Karvonen_NC_2017} MoS$_2$.

The rapid advance of the field has led to the observation of high-harmonic generation (HHG) in ML \ce{MoS2} under intense laser excitation\cite{liu2016high}. However, there are several aspects of the NLO response of TMDs that remain unexplored. For example, nonlinear optics with other TMDs [e.g., third-harmonic generation (THG) in WSe$_2$, MoSe$_2$, and WS$_2$] has not been fully studied \cite{Autere_AM_18}. Furthermore, there is a large deviation in reported experimental values of $\chi^{(2)}$ for 2D materials (including TMDs), which could be partially attributed to differences between measurement conditions (e.g., excitation conditions, sample doping and strain effects) in those studies and also to the different substrates used in the measurements (e.g., different thicknesses and compositions). In fact, especially in the case of 2D materials, the substrate has a significant impact on harmonic generation\cite{Merano_OL_2016,Clark_APL_2015,Miao_AMI_2017}, which makes the comparison of $\chi^{(2)}$ and \textit{$\chi^{(3)}$} values from different measurements problematic. For example in \ce{MoS2}, the reported values of $|\chi^{(2)}|$ at ~800 nm excitation vary from $10^{-7}$ to $10^{-10}$ m V$^{-1}$ (i.e., by three orders of magnitude)\cite{malard2013,li2013,kumar2013second}. Furthermore, different experimental methods, such as two-wave mixing\cite{torres2016third}, multiphoton microscopy\cite{antti2016,woodward2016characterization,autere2017rapid}, and spatial self-phase modulation\cite{wang2016coherent,wang2015tunable}, have been used to quantify the nonlinear susceptibilities of different materials, thus making the comparison even more involved. In conclusion, despite the importance of accurately assessing the NLO susceptibilities of TMDs and shedding light on their dependence on environmental conditions (e.g., as a tool to modulate the response at will), experimental studies so far available are fragmented and do not allow us to establish a systematic comparison between different materials.

Here, we present an experimental study of the second- and third-order NLO properties of group-VI TMDs that is immune to differences in sample and excitation conditions. Monolayers of the four TMD materials MoS$_2$, MoSe$_2$, WS$_2$, and WSe$_2$ are mechanically exfoliated and transferred onto a substrate, in close proximity to each other using a state-of-the-art dry-transfer technique. The effective NLO susceptibilities are then simultaneously determined for all four materials from a single set of SHG and THG images. As a result, the effective bulklike second- and third-order nonlinear susceptibilities of all four materials are directly comparable. The excitation  light source in our experiments is a mode-locked erbium-doped fiber laser with 1560 nm center wavelength. Thus, the resulting SHG and THG signals lie in the visible-to-near-infrared range. Additionally, this provides information about $\chi^{(2)}$ and $\chi^{(3)}$ at 1560 nm, which is important for telecommunication applications.  Moreover, the NLO responses of all four TMDs under consideration are examined with linear and elliptical polarization in this work. Finally, we theoretically calculate their second- and third-order nonlinear susceptibilities through a perturbative expansion of the two-band ${\bf k}\cdot{\bf p}$ Hamiltonian for such media, including the effect of the exciton resonance. These theoretical results are in qualitative agreement with measurements.

\begin{figure}[!htbp]
	\centering
	\includegraphics[width=\columnwidth]{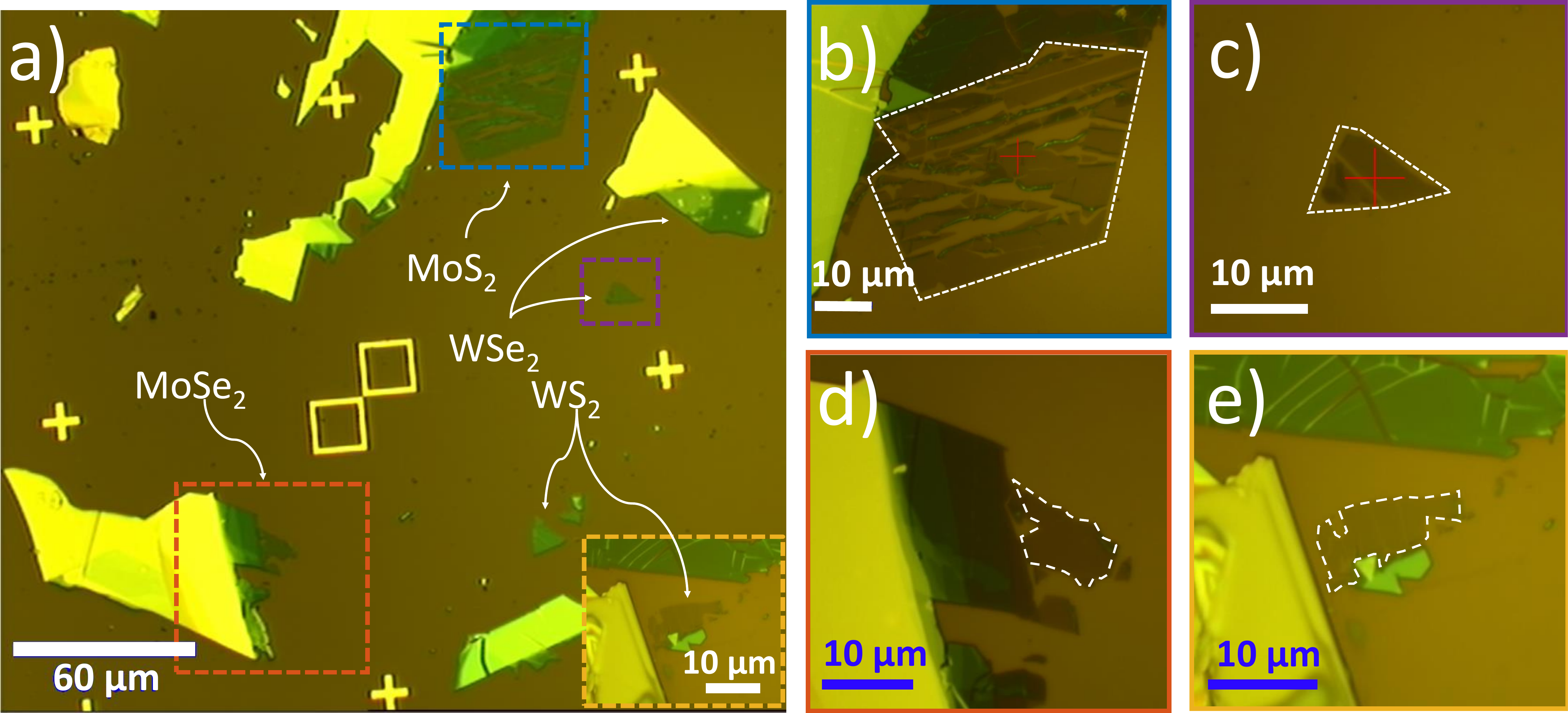}
	\caption{(a) Optical image of four different TMDs positioned close to each other. Zoomed optical images of areas marked in (a) with colored rectangles for (b) MoS$_2$, (c) WSe$_2$, (d) MoSe$_2$, and (e) WS$_2$. White dashed contours in (b)-(e) indicate the monolayer areas.}
	\label{fig: setup}
\end{figure}

\section{Experimental study}

In our experiments, micromechanically exfoliated samples of MoS$_2$, MoSe$_2$, WS$_2$ and WSe$_2$ were transferred onto a Si/SiO$_2$ (285 nm) substrate, close to each other. The sample fabrication process is similar to that reported in Refs. \cite{castellanos2014deterministic,li2015polarization} (fabrication details in the Supplemental Material (SM) \cite{sm}). The transferred flakes were then identified and characterized through optical contrast, Raman spectroscopy and photoluminescence (PL) measurements. The Raman peak separation can be used to extract the layer thickness \cite{li2012bulk,berkdemir2013identification}. Details of these measurements are provided in the SM \cite{sm}. Figure \ref{fig: setup} shows an optical image of the fabricated sample. The ML flake of WS$_2$ is shown as an inset because it is placed slightly farther away from the other materials, outside the image field of view. Additionally, graphene monolayers were exfoliated on a similar substrate for comparison. Strain and doping can have a significant effect on the (nonlinear) optical properties of 2D materials. The possible presence of strain is excluded based on the measured Raman spectra. For example, in Ref. \cite{conley2013bandgap}, it was shown that the E$_{2g}$ peak in \ce{MoS2} shifts by ~4.5 cm$^{-1}$ per percent strain and splits into two separate subpeaks already at ~0.5\% strain, which we do not see in any of our measurements (see Figs. 2 and 3 in the SM). Furthermore, the 2D materials are obtained from undoped bulk crystals and, since we are using a dry-transfer technique, the samples are not chemically doped. This is different from other transfer methods, in which the chemicals used during the transfer process might introduce the doping effect. The doping level can also be modified due to the substrate. For example, \ce{SiO2} can have a high degree of charge impurities at the surface which can lead to an altered doping level of the sample\cite{buscema2014effect}. The effect of doping on the nonlinear optical response of 2D materials is still largely unexplored. However, in Ref. \cite{Clark_APL_2015} SHG from \ce{MoS2} on \ce{SiO2} and polymer substrates was measured. With a fundamental wavelength at 1600 nm they measured a $\chi^{(2)}$ of 6.3 pm/V on the \ce{SiO2} substrate and 7 pm/V on the polymer substrate, suggesting that the doping from the substrate may not be substantial. As noted earlier, to best of our knowledge, there are no studies on the effect of chemical- or substrate-induced doping on nonlinear optical properties of 2D materials at different wavelengths, so it is not possible to accurately assess the different contributions on the nonlinear optical responses of the four TMDs studied here. However, this is a highly interesting topic for future research.

\begin{figure}[h!]
	\centering
	\includegraphics[width=\columnwidth]{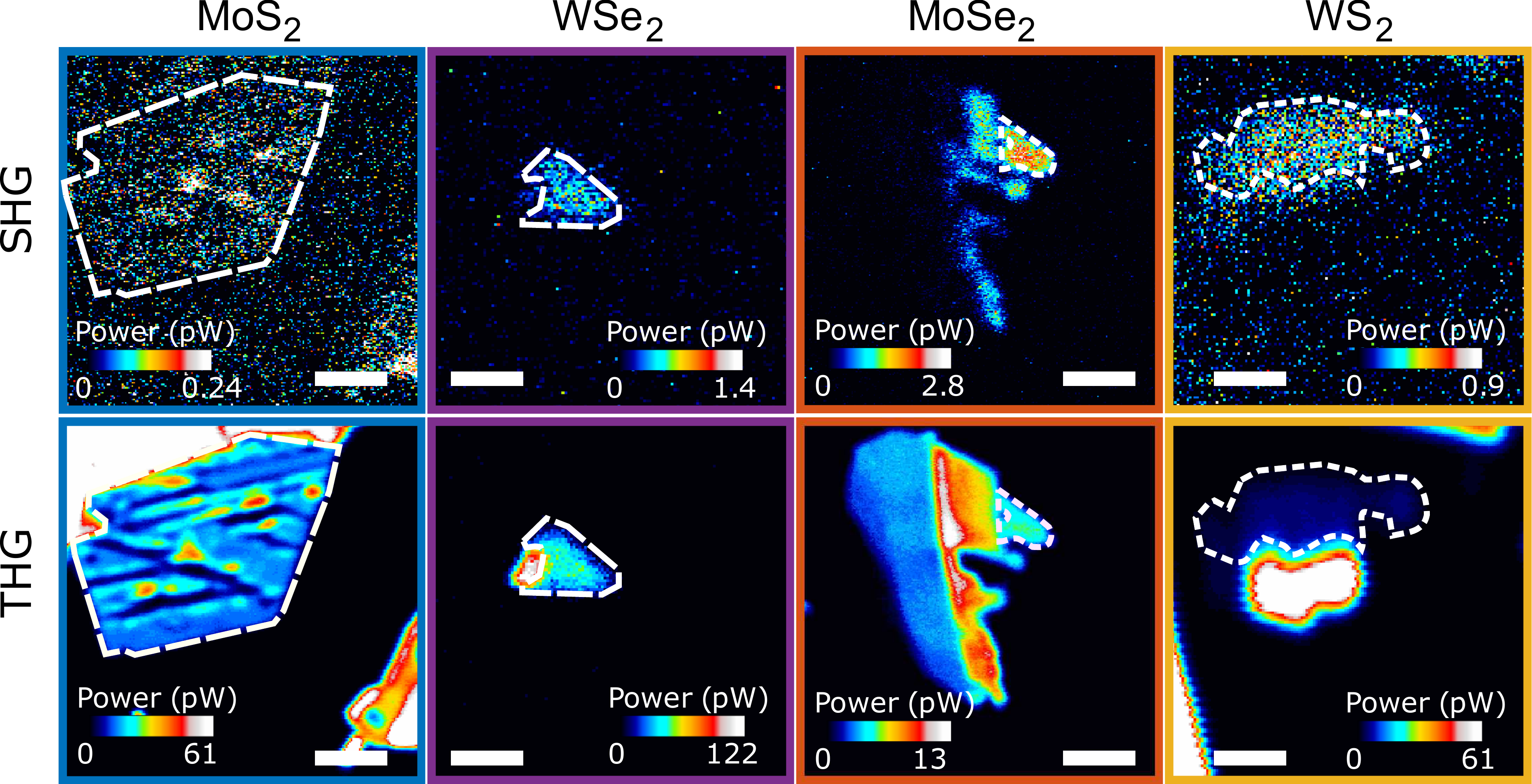}
	\caption{SHG (top) and THG (bottom) images from the areas marked with colored dashed rectangles in the optical image of Fig. \ref{fig: setup}(a). Blue, MoS$_2$; purple, WSe$_2$; orange, MoSe$_2$; yellow, WS$_2$.The scale bar for MoS$_2$, WSe$_2$ and MoSe$_2$ is 10 $\mu$m, and that for WS$_2$ is 5 $\mu$m.}
	\label{fig: SHG THG images}
\end{figure}

The SHG and THG for the four materials under examination (Fig. \ref{fig: SHG THG images})  were collected using a mode-locked fiber laser with 1560~nm center wavelength. Each different area (located within a distance $<150~\mu$m from each other) possesses exfoliated TMD flakes, whose thicknesses range from one to a few atomic monolayers (see the SM for thickness determination). The small distance between the locations of the exfoliated MoS$_2$, MoSe$_2$, WS$_2$, and WSe$_2$ allows us to easily compare the optical nonlinearities of the different materials. As a result, this excludes the effect of substrate and varying measurement conditions on the measured susceptibilities because the recorded SHG and THG powers can be obtained from the same SHG or THG image for all materials. The locations of the nonlinear microscopy images are indicated by dashed rectangles in the optical image in Fig. \ref{fig: setup}. We observed a SHG signal and a strong THG signal from all four TMDs, within a single image, of which Fig.\ref{fig: SHG THG images} selects zoomed regions.

\begin{figure}[h!]
	\centering
	\includegraphics[width=\columnwidth]{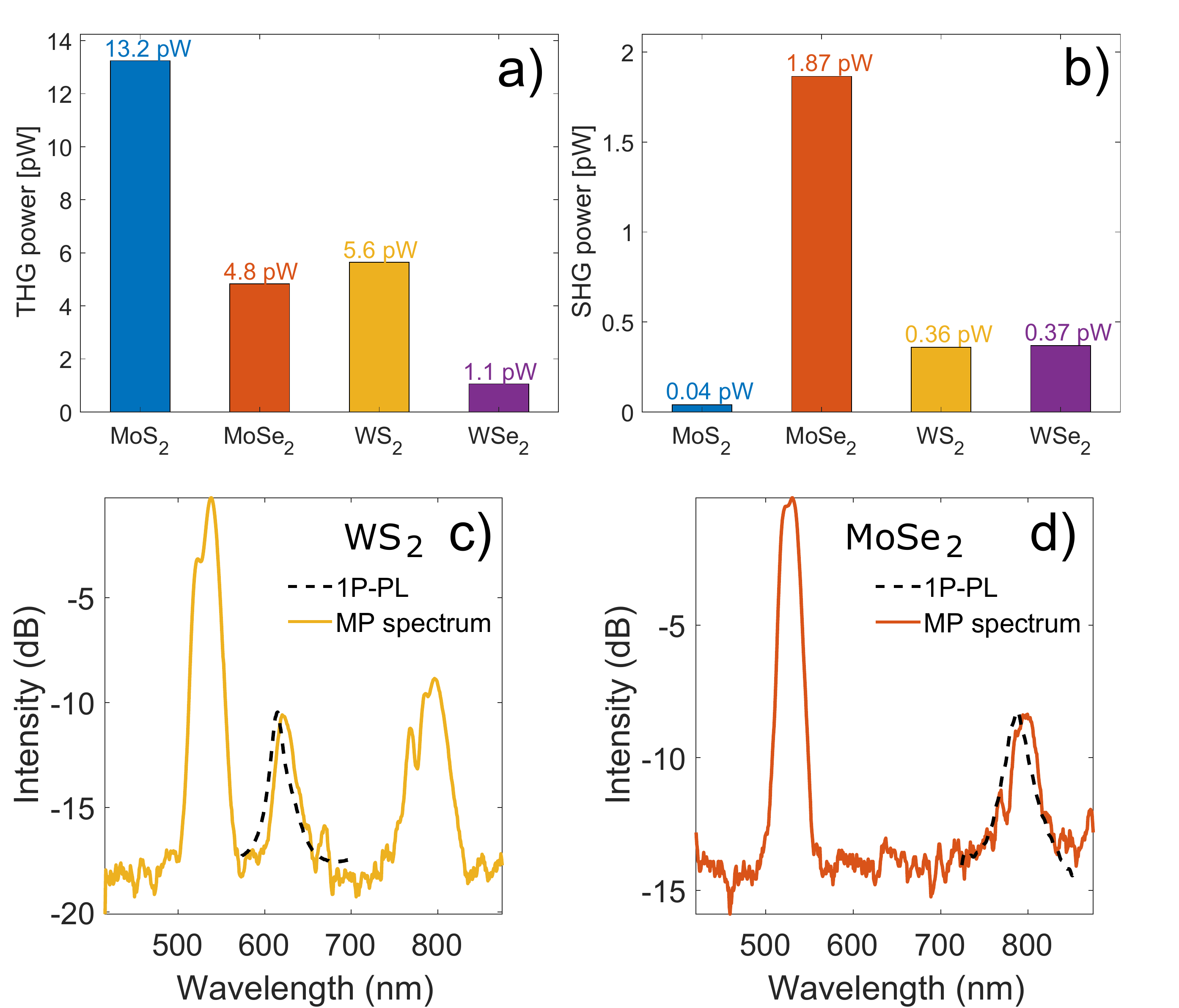}
	\caption{Measured average powers of (a) THG and (b) SHG from monolayers of all four materials with 20 mW pump power ($\sim$2.7 kW peak power). Comparison between one-photon and multiphoton excited spectrum of (c) WS$_2$ and (d) MoSe$_2$. Note that the intensities of one-photon and multiphoton spectra in (c) and (d) are not to scale.}
	\label{fig: PS and spectra}
\end{figure}

The reported values of SHG and THG signals are all obtained from the locations in which each TMD flake is a single layer thick. Prior to comparison, however, in order to prove that the measured signals at 780 and 520 nm in fact originate from SHG and THG, respectively, power dependent nonlinear microscopy measurements were performed for all four materials (see SM, Fig. 10). These measurements clearly indicate $P^2$ (SHG) and $P^3$ (THG) dependencies, with the incident light power $P$. Note that this is the first time that THG can be observed in WSe$_2$, MoSe$_2$, and WS$_2$. The average powers of SHG and THG from monolayers of all four materials are shown in Figs. \ref{fig: PS and spectra}(a) and \ref{fig: PS and spectra}(b). Normally at a reasonable pump power, THG intensity is expected to be lower than SHG intensity, due to the weaker intrinsic response of the higher-order nonlinear processes\cite{Boyd}. It is surprising that THG is clearly stronger than SHG in all four TMDs at such a low average pump power (e.g. Figs. 3(c) and 3(d)). The same effect was recently observed in \ce{MoS2}, and explained with the contribution of trigonal warping to the second-order response\cite{antti2016}. With low incident photon energies (0.8 eV here and in Ref. \citenum{antti2016}), SHG is weaker than expected for \ce{MoS2} due to near-isotropic bands contributing to the SHG signal. Only trigonal warping breaks the approximate rotational invariance of the monolayer \ce{MoS2} band structure, causing the SHG emission\cite{antti2016}. More insight into the possible effects of trigonal warping or other causes (e.g. excitons) leading to the large observed ratio between THG and SHG, could be obtained by measuring the SHG and THG from all four materials with a large range of excitation wavelengths. Note that the SHG intensity from \ce{MoSe2} is lower than the intensity of THG, even though the SHG is on resonance with the \textit{A} exciton\cite{le2016nonlinear,malard2013,wang2015tunable,Seyler2015,Wang2015exciton}. Furthermore, we observe clearly distinct THG and SHG signals for different TMDs. For instance, THG is largest from MoS$_2$ and smallest from WSe$_2$. In contrast, SHG from MoSe$_2$ is $\sim$ 4--40 times larger than that from the other materials. This can be attributed to resonant enhancement in MoSe$_2$ because the energy of the \textit{A} exciton in this material ($\sim$ 1.57~eV, 790~nm\cite{tonndorf13}) matches well with the wavelength (780~nm, 1.59~eV). In fact, the spectral overlap of excitonic PL and SHG is well visualized in Fig. \ref{fig: PS and spectra}(d), which shows the PL spectrum measured with 532 nm excitation, and the multiphoton (MP) excited spectrum [containing SHG, THG, and two-photon excited luminescence (2PL)] for MoSe$_2$.

We note that the MP excited spectrum of \ce{WS2} also shows a peak at $\sim$ 615 nm (2.01 eV), corresponding to the location of the PL peak [Fig. \ref{fig: PS and spectra}(c)]. Thus, we attribute this peak to three-photon excited luminescence (3PL) from monolayer \ce{WS2}. Since 3PL ensues from a fifth-order NLO process\cite{Qian_AM_2015}, the probability of 3PL occurrence is very low. Interestingly, the intensity of 3PL is in the same range as the intensity of SHG. 2PL spectroscopy has been used to study the excitons in TMD monolayers because with 2PL it is possible to probe dark exciton states, which are forbidden by selection rules for one-photon excitation\cite{ye2014probing,he2014tightly}. Graphene and \ce{MoS2} quantum dots have been used as 2PL probes in cellular and deep-tissue imaging\cite{liu2013strong,dai2015tunable}. However, 3PL can provide better spatial resolution\cite{yu2013high} and enable alternative excitation wavelengths, thus \ce{WS2} might find new applications in biomedical imaging. Furthermore, 3PL spectroscopy might provide an alternative method for probing excitonic features in monolayer TMDs. \\
We estimate the effective second- and third-order nonlinear susceptibilities $|\chi^{(2)}_{\text{eff}}|$ and $|\chi^{(3)}_{\text{eff}}|$ of all TMDs from the measured average SHG and THG powers. The sheet susceptibility values, $\chi^{(n)}_s$, are estimated with the methods described in Ref. \citenum{woodward2016characterization}, by fitting the measured average powers to the following two equations
\begin{align}
P_{\text{SHG}}&=\frac{16\sqrt{2}S|\chi_s^{(2)}|^2\omega^2}{c^3\varepsilon_0 f \pi r^2 \tau (1+n_2)^6}P_1^2 \label{eq:chi2},\\
P_{\text{THG}}&=\frac{64\sqrt{3}S^2|\chi_s^{(3)}|^2\omega^2}{c^4\varepsilon_0^2 (f \pi r^2 \tau)^2 (1+n_2)^8}P_1^3,\label{eq:chi3}
\end{align}
where $S=0.94$ is the shape factor for Gaussian pulses, $\tau$ is the temporal pulse width, $P_{1}$ is the incident average power of the pump beam, $f$ is the repetition rate, $n_2$ is the refractive index of the substrate at the pump wavelength, and $\omega$ is the angular frequency of the pump. The effective bulklike second-order susceptibility of TMDs is obtained from the sheet susceptibilities as $$|\chi_{\text{eff}}^{(n)}|=\frac{|\chi_s^{(n)}|}{t},$$ where $t$ is the thickness of the TMD monolayer, 0.65 nm.
\begin{figure}[h!]
	\centering
	\includegraphics[width=\columnwidth]{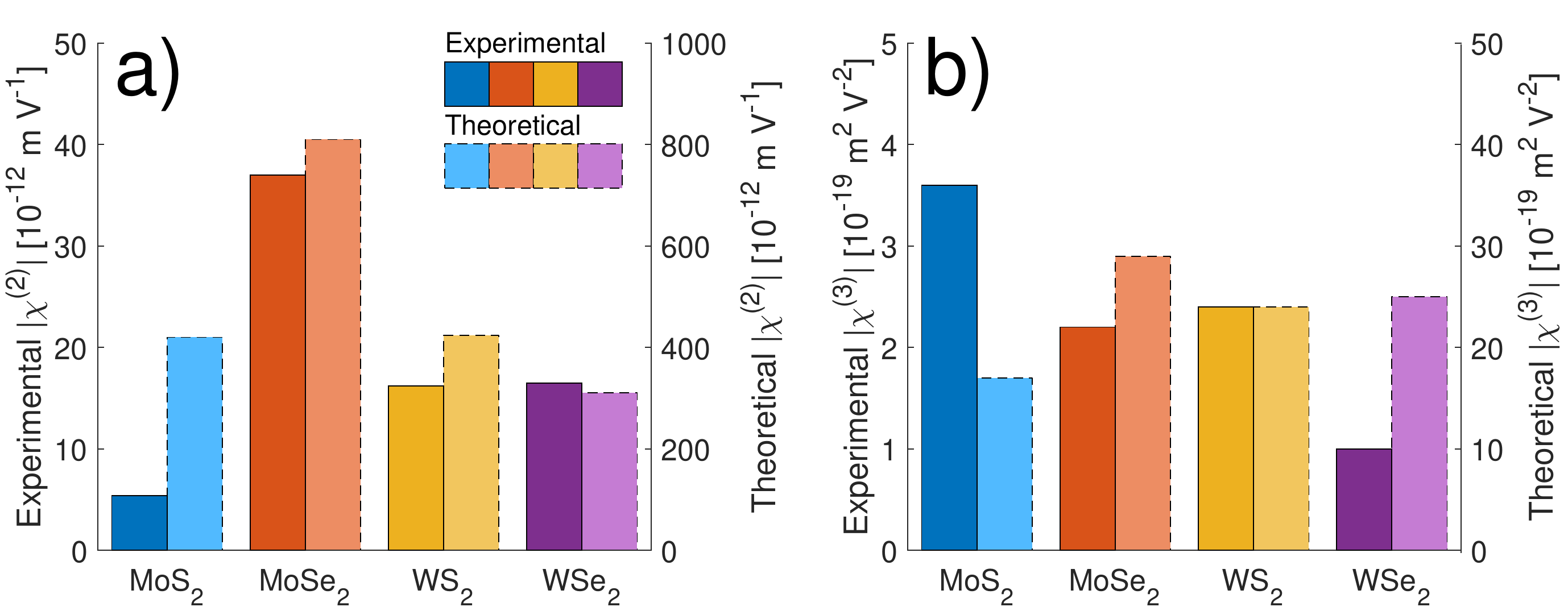}

	\caption{Comparison of experimental and theoretical (a) $|\chi^{(2)}|$ and (b) $|\chi^{(3)}|$ of four TMDs at 1560 nm excitation.}
	\label{fig: comparison}
\end{figure}

The $|\chi^{(2)}_{\text{eff}}|$ and $|\chi^{(3)}_{\text{eff}}|$ values measured from different TMDs in this work are presented in Table 1 and Fig. 4.
Furthermore, $|\chi^{(2)}_{\text{eff}}|$ and $|\chi^{(3)}_{\text{eff}}|$ values for various monolayer TMDs from other measurements reported in the literature are presented in Table 1 in the SM.

Note that $|\chi_{\text{eff}}^{(2)}|$ values obtained in this work lie in the range between \smos~$\times~10^{-12}$ and \smose~$\times~10^{-12}$ ${\mathrm{m}}\mathrm{V}^{-1}$ for all ML-TMDs. These values are in good agreement with those reported in the literature for TMDs when they have been measured with excitation wavelength in the IR region. For instance, the literature values of $|\chi_{\text{eff}}^{(2)}|$ for MoS$_2$\cite{le2016impact,woodward2016characterization,antti2016,clark2014strong} ranges between $2.2~\times10^{-12}$ and $29~\times10^{-12}$ ${\mathrm{m}}\mathrm{V}^{-1}$ and thus match reasonably well with the value of \smos~$\times10^{-12}$ ${\mathrm{m}}\mathrm{V}^{-1}$ obtained in this work for 1560 nm excitation.

We note that the $|\chi_{\text{eff}}^{(2)}|$ for MoS$_2$ is two orders of magnitude smaller than what has been measured with 800 nm excitation\cite{li2013,malard2013}. On the other hand, $|\chi_{\text{eff}}^{(3)}|$ values for all characterized TMDs are in the range between \twse~$\times10^{-19}$ ${\mathrm{m^2}}\mathrm{V}^{-2}$ and \tmos~$\times10^{-19}$ ${\mathrm{m^2}}\mathrm{V}^{-2}$. We also find excellent agreement with previous literature values for $|\chi^{(3)}_{\text{eff}}|$ when measured at a similar wavelength. For instance, the magnitude of the $|\chi^{(3)}_{\text{eff}}|$ values for MoS$_2$ is of the order of $1~\times10^{-19}$ ${\mathrm{m^2}}\mathrm{V}^{-2}$ (see Refs. \cite{woodward2016characterization,antti2016,wang2013third}) and therefore in the same range as the value of \tmos~$\times10^{-19}$ ${\mathrm{m^2}}\mathrm{V}^{-2}$ reported in this work. The effect of the substrate should also be taken into account when comparing the values. In Ref. \citenum{woodward2016characterization} the bulklike $|\chi^{(2)}|$ and $|\chi^{(3)}|$ of $\ce{MoS2}$ were measured on glass and on Si/$\ce{SiO2}$ substrates. It was found that the $|\chi^{(2)}|$ did not exhibit significant change but the $|\chi^{(3)}|$ was enhanced by a factor of 5 due to the interferometric effect caused by the multilayer structure. However, this does not affect the comparison between the four materials, as the effect holds for all of them, in this experiment.\\

\begin{figure}[ht!]
	\centering
	\includegraphics[width=\columnwidth]{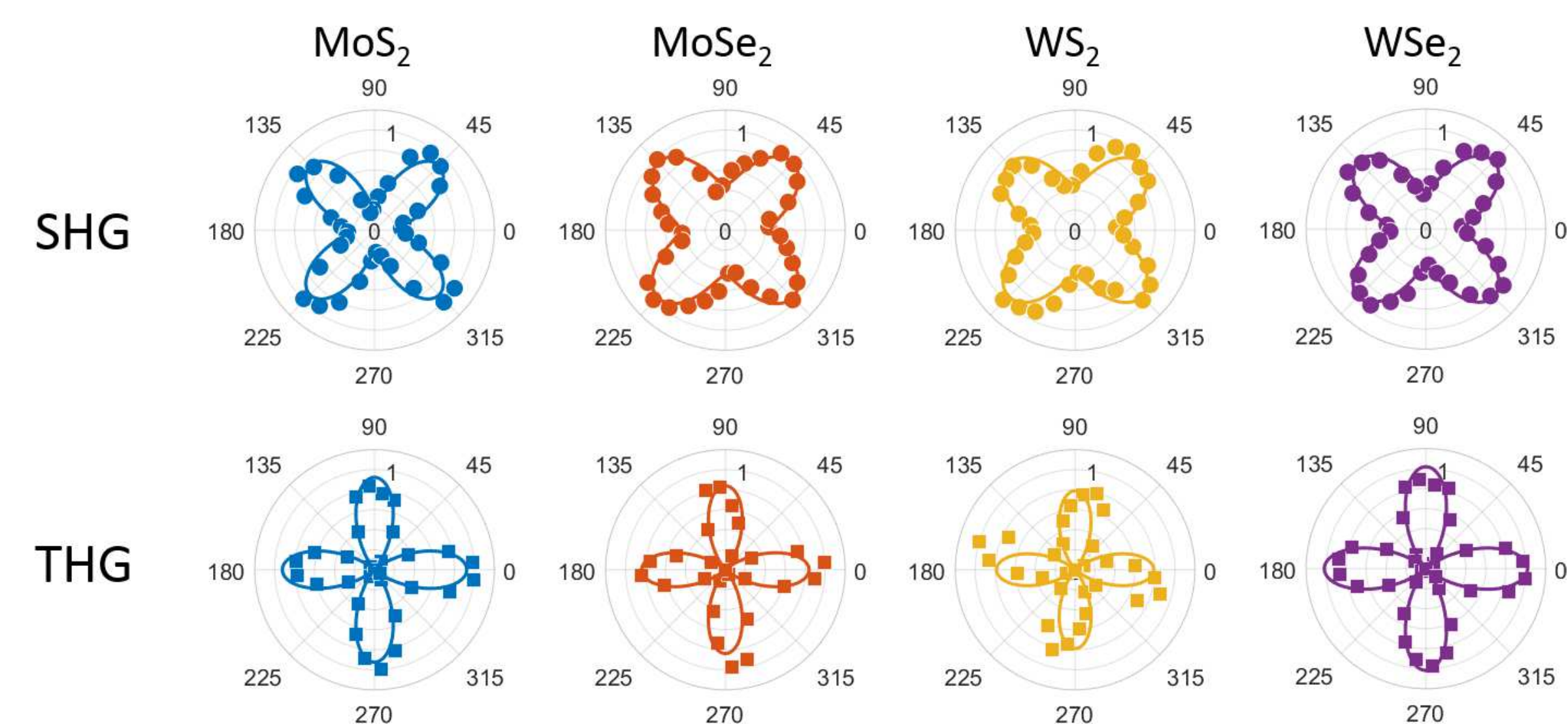}
	\caption{Polar plots of normalized SHG and THG signals as a function of the quarter-wave plate (QWP) angle. The angle is defined between the polarization of the incident laser and the fast axis of the QWP.}
	\label{fig: polar images}
\end{figure}

The intensities of SHG and THG depend strongly on the polarization state of the pump and the crystallographic orientation of the sample\cite{antti2016,li2013,liu2016high,woodward2016characterization}. In order to explore this effect, we have measured SHG and THG from all four materials using elliptically polarized excitation light with varying degree of ellipticity. Figure \ref{fig: polar images} shows the measured SHG and THG intensities as a function of incident light polarization state. As shown in Fig. \ref{fig: polar images}, the THG signal is strongest for linearly polarized excitation light and smallest for circular polarization. In contrast, the SHG signal is strongest for circularly polarized excitation light and smallest for linearly polarized excitation light.  Because all four materials belong to the point group $D^1_{\text{3h}}$, similar results are obtained for the other crystals, MoSe$_2$, WS$_2$, and WSe$_2$. This paves the way for tailoring the nonlinear optical properties of 2D materials arranged in heterostructures.

\section{Theoretical analysis and comparison to experimental values}
In order to obtain further insight into the origin of the NLO behavior under study, we theoretically calculate the second- and third-order nonlinear susceptibilities of all four materials through a perturbative expansion of the two-band ${\bf k}\cdot{\bf p}$ Hamiltonian for such media \cite{LWY2013} under the minimum coupling prescription $\mbox{\boldmath${\pi}$} = {\bf p} + e {\bf A}(t)$, where $e$ is the electron charge, ${\bf A}(t)$ is the potential vector of the impinging light beam, and ${\bf p}$ and $\mbox{\boldmath${\pi}$}$ are the electron momentum and quasimomentum, respectively.
\begin{table}[h!]
	\small
	\caption{ Theoretical and experimental $|\chi^{(2)}|$ and $|\chi^{(3)}|$ values of different TMD materials.}
	\label{tab: Theor x2}
	\begin{ruledtabular}
		\begin{tabular}{lllll}
			& \multicolumn{2}{c}{ $|\chi^{(2)}|$ $\left[10^{-12}\frac{\mathrm{m}}{\mathrm{V}}\right]$} & \multicolumn{2}{c}{$ |\chi^{(3)}|$ $\left[10^{-19}\frac{\mathrm{m}^2}{\mathrm{V}^2}\right]$} \\
			Material &Theor.&Expt.&Theor.&Expt.\\
			\hline
			\ce{MoS2}&  420 & \smos& 17  &\tmos\\
			\ce{MoSe2}& 810 & \smose&  29 &\tmose\\
			\ce{WS2}&   424 & \sws&  24 &\tws\\
			\ce{WSe2}&  311 & \swse&  25 &\twse\\
		\end{tabular}
	\end{ruledtabular}
\end{table}
The two-band ${\bf k}\cdot{\bf p}$ Hamiltonian is obtained by fitting the valence and conduction bands of the tight-binding Hamiltonian reported in the literature \cite{LWY2013} that account for both nondegenerate valleys and spin-orbit coupling. The effect of the exciton resonance on the nonlinear parameters of MoSe$_2$ is taken into account by introducing an effective exciton energy level in the single-particle Hamiltonian after fitting the linear conductivity with fully numerical Bethe-Salpeter calculations \cite{YAMBO}. Results are presented in Table \ref{tab: Theor x2} and compared with experiment in Fig. \ref{fig: comparison}, while technical details of the calculations are provided in the SM \cite{sm}. These theoretical calculations predict a generally higher value of the nonlinear coefficients, roughly one order of magnitude larger than the corresponding measured quantities. Such calculations confirm that MoSe$_2$ is the ML TMD with highest $|\chi^{(2)}|$ nonlinear coefficient, although the relative difference with respect to other TMDs is not as marked as in experiments. A mismatch of relative values across different materials also appears in the $\chi^{(3)}$ calculations, which predict that MoSe$_2$ has the highest $|\chi^{(3)}|$, in contrast to experiment, in which MoS$_2$ exhibits the highest third-order nonlinearity. We envisage that this may be due to the effect of the substrate on electron many-body dynamics, which deserves further attention. In addition, the substrate is expected to induce subtle modifications on the band structure. While the linear response of the 2D layers under consideration remains unaffected because it mainly depends on the energy bandgap, the NLO response originates in the anharmonicity of the bands and thus it is much more sensitive to small modifications in the band structure arising from the interaction of ML-TMDs with the substrate. Comparison with our experimental results provides an indication of the trends when examining different materials and also on the orders of the magnitude of the effects. Nevertheless, future theoretical efforts beyond the scope of this work are required to obtain a good quantitative agreement. We envisage that, in order to improve predictions, future theoretical efforts should include more electronic bands, account for the interaction of the considered two-dimensional media with the substrate (which modifies the electronic band structure), and ultimately rely on first-principle simulations to fully describe the nonlinear exciton dynamics beyond the effective exciton band model used in our calculations (see the SM \cite{sm}).

\section{Conclusions}

In summary, in this work we demonstrate third-harmonic generation in WSe$_2$, MoSe$_2$, and WS$_2$, and three-photon photoluminescence in TMDs. We also report the first direct comparison of second- and third-order optical nonlinearities in MoS$_2$, MoSe$_2$, WS$_2$, and WSe$_2$.
The $\chi^{(2)}$ of MoSe$_2$ is found to be approximately two to six times larger than that of the other TMDs examined here.
We attribute this effect to resonant enhancement of SHG in MoSe$_2$. The third-order nonlinear susceptibility $\chi^{(3)}$ of all four materials was found to be comparable to that of graphene, with the largest value $|\chi^{(3)}_{\text{eff}}| = \tmos\times10^{-19} {\mathrm{m^2}}\mathrm{V}^{-2}$ observed for MoS$_2$. We obtain further insight into the NLO properties by theoretically calculating the second- and third-order nonlinear susceptibilities of all four materials, in qualitative agreement with measurements.

Furthermore, the effect of the degree of elliptical polarization of the incident light on the SHG and THG signals was examined and we found that the SHG signal was enhanced and the THG one was completely suppressed with circular polarization. Experimental results fit very well with expected values based on previously reported expressions derived from the crystal symmetry for MoS$_2$. The results presented here provide valuable information about the nonlinear properties of the different TMDs for the design of devices based on 2D materials and their heterostructures in a wide range of applications, such as on-chip light sources and all-optical signal processing.
\begin{acknowledgements}
	We acknowledge funding from the Academy of Finland (No. 276376, No. 284548, No. 295777, No. 298297, No. 304666, No. 312551, No. 312297, and No. 314810), TEKES (NP-Nano, OPEC), the Nokia Foundation, Tekniikan edist\"amiss\"a\"ati\"o (TES), Aalto Centre of Quantum Engineering, the China Scholarship Council, Spanish MINECO (MAT2017-88492-R and SEV2015-0522),the European Commission (REA Grant No. 631610, Graphene Flagship 696656) and AGAUR (FI$\_$B-00492-2015 and 2017-SGR-1651). We acknowledge the provision of facilities and technical support by Aalto University at Micronova Nanofabrication Centre.
\end{acknowledgements}

\bibliographystyle{apsrev4-1}

\end{document}